\documentclass[aps,prl,twocolumn,longbibliography]{revtex4-2}

\usepackage{amssymb}
\usepackage{amsbsy}
\usepackage{amsmath}
\usepackage{graphicx}
\usepackage{graphics}
\usepackage{setspace}
\usepackage{array}
\usepackage{color}
\usepackage{fontenc}
\usepackage{textcomp}
\usepackage{bm}
\usepackage{float}
\usepackage[bookmarks=false,linkcolor=blue,urlcolor=blue,colorlinks,citecolor=blue]{hyperref}

\DeclareMathOperator{\re}{Re}

\newcommand{\vex}[1]{\bm{\mathrm{#1}}}

\newcommand{\blue}[1]{{\color{blue}{#1}}}

\newcommand{\bsub}{\begin{subequations}}
\newcommand{\esub}{\end{subequations}}

\graphicspath{{../Figures/}}

\begin{document}
\title{Multiphoton Fingerprints of Altermagnetic Spin Splittings}
\author{Sayed Ali Akbar Ghorashi}
\email{ghorashi@sas.upenn.edu}
\author{Andrew M. Rappe}
\affiliation{$^1$Department of Chemistry, University of Pennsylvania, PA, USA}

\date{\today}

\newcommand{\be}{\begin{equation}}
\newcommand{\ee}{\end{equation}}
\newcommand{\bea}{\begin{eqnarray}}
\newcommand{\eea}{\end{eqnarray}}
\newcommand{\h}{\hspace{0.30 cm}}
\newcommand{\vs}{\vspace{0.30 cm}}
\newcommand{\n}{\nonumber}

\begin{abstract}
We systematically investigate multiphoton absorption as a polarization-resolved nonlinear optical probe of planar altermagnets (ALMs). We show that the angular harmonic of the altermagnetic spin splitting fixes the lowest optical absorption at which a symmetry-selective response appears: two-photon absorption for $d$-wave order, four-photon absorption for $g$-wave order, and six-photon
absorption for $i$-wave order. In each case, there exists a polarization channel locked to the symmetry harmonic of the altermagnetic texture in which the direct $n$-photon contribution to the transition matrix element is absent. This changes the frequency scaling of the absorption rate relative to other polarization channels and provides a direct optical fingerprint of the underlying altermagnetic harmonic. Our results establish a hierarchy of nonlinear spectroscopic signatures that distinguishes $d$-, $g$-, and $i$-wave altermagnetic spin splittings beyond linear response.
\end{abstract}
\maketitle

\begin{figure}
    \centering
    \includegraphics[width=1\linewidth]{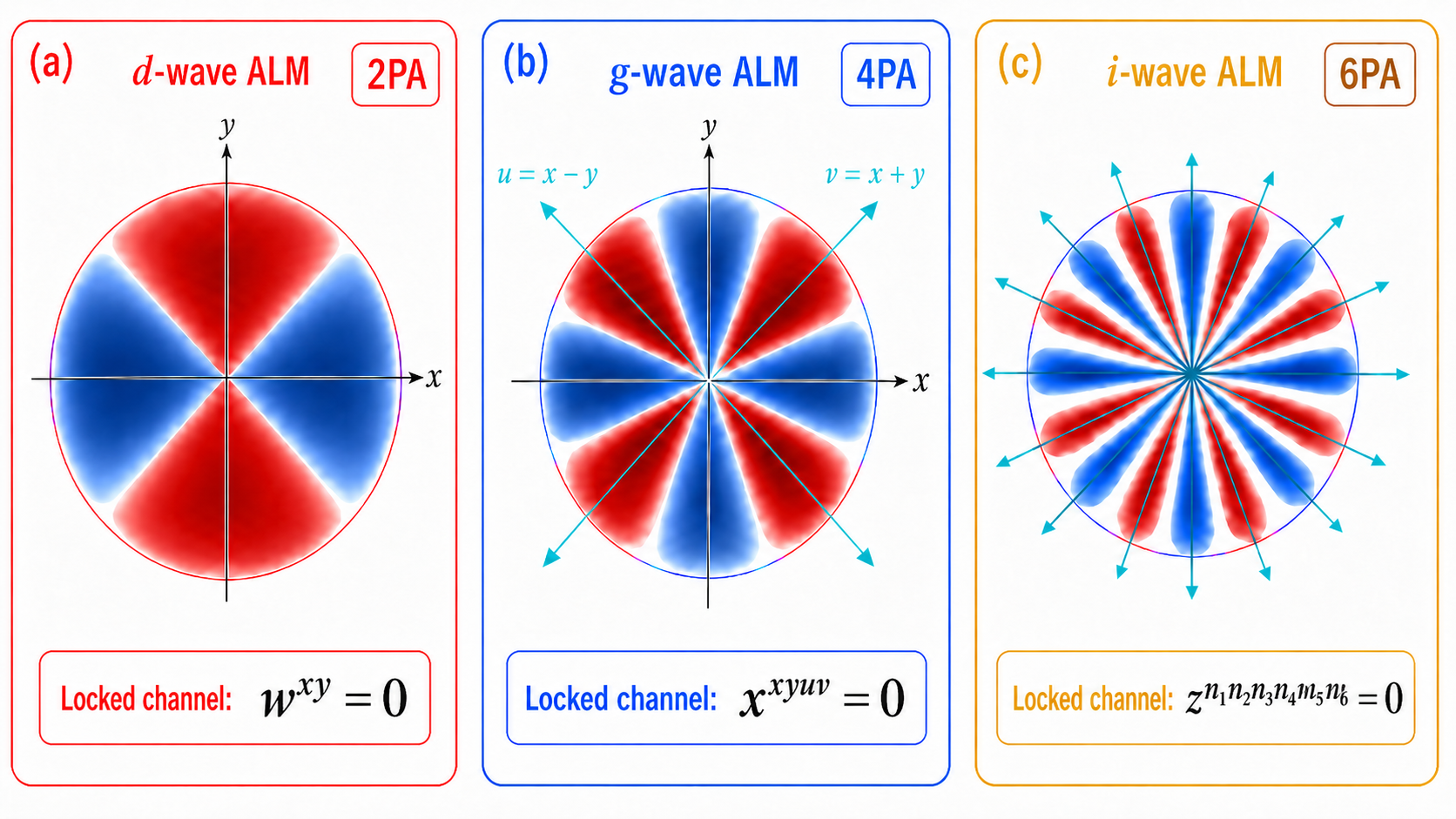}
    \caption{The altermagnetic harmonic determines the lowest absorption order
with a symmetry-selective response: 2PA for $d$-wave, 4PA for
$g$-wave, and 6PA for $i$-wave order.}
    \label{fig:adpic}
\end{figure}

\blue{\emph{Introduction}}.--- Altermagnets, characterized by their unique collinear magnetic order and net zero magnetization, present intriguing opportunities due to their distinctive spin-split band structure governed by crystal symmetries rather than relativistic spin–orbit coupling alone. \cite{altermagnet1,altermagnet2,altermagnetismreview,song2025altermagnets,jungwirth2025altermagnetism,ghorashiALMTSC,fernandes2023topological,PhysRevLett.132.176702,TSCALM2,m8lc-f8gk,jiang2025altermagnetisminducedsurfacechern,tm58-lbdl,PhysRevLett.134.106801,PhysRevLett.134.106802}
The development of characterization schemes capable of revealing definitive hallmarks of altermagnetism is an active and highly sought-after direction in current research \cite{jungwirth2026symmetry,PhysRevB.111.184408,PhysRevMaterials.8.L041402,ghorashiALMnonlinear,gly7-jzfl,sunko2026linear,PhysRevB.108.L100402,das2026linear}. This need is particularly acute, because several reported signatures, including the anomalous Hall effect, are constrained by symmetry and therefore cannot serve as unambiguous hallmarks of altermagnetism \cite{PhysRevLett.133.206401}. \\
\indent Optical probes \cite{gly7-jzfl,gray2024time,bzzy-ngcs} are, in principle, well suited to access band structure and symmetry properties, but linear-response techniques face intrinsic limitations in altermagnets. One-photon absorption and related linear optical conductivities couple to the electronic structure only through a single light--matter vertex. By contrast, altermagnetic spin splitting is encoded in higher-order momentum harmonics of the Bloch Hamiltonian. As a result, linear optical responses are generally not sufficient to fully capture the symmetry character of altermagnetic order.\\
\indent In this work, we demonstrate that multiphoton absorption (MPA) provides a particularly appealing nonlinear mechanism in this context \cite{nathan1985review,hayat2011applications}. In an $n$-photon absorption process, electronic transitions are mediated by virtual intermediate states and described by higher-order perturbation theory. Crucially, multiphoton processes relax the parity and angular-momentum selection rules that constrain single-photon absorption, enabling access to otherwise forbidden interband transitions. They are also highly sensitive to the polarization state and relative phase of the incident fields, allowing interference between distinct excitation pathways.\\
\indent Motivated by these considerations, we explore MPA as a symmetry-resolved probe of altermagnetic order. By analyzing the structure of the nonlinear optical transition amplitudes in representative altermagnetic systems, We show that the angular harmonic of the altermagnetic spin splitting determines the lowest absorption order at which a symmetry-selective response emerges: two-photon absorption for $d$-wave order,
four-photon absorption for $g$-wave order, and six-photon absorption
for $i$-wave order Fig.~\ref{fig:adpic}. In each case, a polarization channel locked to the
symmetry of the altermagnetic texture nullifies the direct $n$-photon
contact term. As a result, specific symmetry guided components of MPA possess distinct frequency scaling. This shows that characteristic polarization and frequency dependence are directly related to the underlying spin–crystal symmetry. These findings establish multiphoton spectroscopy as a viable route to detect and characterize altermagnetism beyond the constraints of linear-response techniques.

\blue{\emph{Formalism}}.---Let us start with a general two-band Hamiltonian for altermagnets with an arbitrary order parameter $J(\vex{k})$,
\begin{align}
    H&=t\left(\cos(k_x)+\cos(k_y)\right)+\lambda\left(\sin(k_x)\sigma^y-\sin(k_y)\sigma^x\right)\cr
    &+J(\vex{k})\sigma^z
\end{align}
where $t$, $\lambda$ denote the hopping and spin-orbit coupling (SOC) strength.
The multiphoton absorption coefficient is obtained from $n$th-order time-dependent perturbation theory, following the framework introduced by G"oppert-Mayer for simultaneous photon absorption~\cite{nathan1985review}. For a crystalline material, the same perturbative structure describes a direct vertical transition from an initial valence band $v$ to a final conduction band $c$, mediated by the coherent absorption of $n$ photons of frequency $\omega$. The resulting $n$-photon transition probability is,

\begin{align}
    W_n=\frac{2\pi}{\hbar}\int_{BZ}|\mathcal{M}|^2\delta[E_c-E_v-n\hbar\omega]\frac{d^2\mathbf{k}}{(2\pi)^2}
\end{align}
where
\begin{align}\label{MPA1}
    \mathcal{M}=\sum_{m,l,\ldots,i}\frac{\langle\psi_c|H^{\mathrm{int}}|\psi_m\rangle\langle\psi_m|H^{\mathrm{int}}|\psi_l\rangle}{E_m-E_l-(n-1)\hbar\omega}\ldots\frac{\langle\psi_i|H^{\mathrm{int}}|\psi_v\rangle}{E_i-E_v-\hbar\omega}
\end{align}
and $W_n$ is the $n$th-order transition probability rate per unit volume. $\psi_i,\psi_j,\ldots$ are the Bloch functions of the crystalline electrons in bands $i,j,\ldots$ with energies $E_i,E_j,\ldots$, respectively. Energy conservation is enforced by the delta function, and the summation is taken over all possible intermediate states for a given transition. The integration over $\mathbf{k}$ is carried out over the entire first Brillouin zone. The light--matter interaction is taken in the dipole approximation as $H^{\mathrm{int}}=-e\,\mathbf{A}\cdot\mathbf{v}$. The multi-photon absorption coefficient is then related to $W_n$ by
\begin{align}
    \alpha_n=\frac{W_n n\hbar\omega}{I^n},
\end{align}
where $I$ is the incident radiation intensity.

While Eq.~\ref{MPA1} captures the standard sequential (paramagnetic) processes, it is not, by itself, sufficient to yield a physically consistent result in crystalline systems. In the velocity gauge, the light--matter coupling originates from the minimal substitution $\mathbf{k}\to\mathbf{k}+\mathbf{A}$, under which the Hamiltonian expands as
\begin{align}
    H(\mathbf{k}+\mathbf{A}) = H(\mathbf{k}) + A_a\,\partial_a H + \frac{1}{2}A_a A_b\,\partial_a\partial_b H + \cdots.
\end{align}
The first-order term reproduces the interaction $-e\,\mathbf{A}\cdot\mathbf{v}$ used above, while the second-order term gives rise to an additional two-photon vertex proportional to $\partial_a\partial_b H$. This contribution, commonly referred to as the contact (or diamagnetic) term, does not appear explicitly in the Göppert-Mayer expression because Eq.~\ref{MPA1} is written in terms of successive single-photon interactions only. However, in a band structure context, retaining only the sequential terms leads to gauge-dependent results and spurious divergences associated with the energy denominators. The contact term contributes at the same order in the field and exactly compensates these unphysical contributions, ensuring that the total transition amplitude remains finite and gauge invariant. Therefore, for a faithful description of multi-photon absorption in solids, the full amplitude must be understood as containing both the sequential processes of Eq.~\ref{MPA1} and the additional contact contribution arising from the second-order expansion of the Hamiltonian.

\blue{\emph{2PA in $d$-wave ALM}}.---Restricting to the degenerate two-photon case ($n=2$), the transition amplitude obtained from Eq.~(2) takes the form
\begin{align}
    \mathcal{M}^{ab}
    =
    \sum_{m}
    \left[
    \frac{\langle \psi_c | v^a | \psi_m \rangle \langle \psi_m | v^b | \psi_v \rangle}{E_m - E_v - \hbar \omega}
    +
    (a \leftrightarrow b)
    \right].
\end{align}
As noted above, this expression contains only the sequential contribution from two successive single-photon processes. To avoid an incomplete velocity-gauge response, we therefore also include the direct contact term at the same order in the field.

For a two-band system, the summation over intermediate states can be evaluated explicitly by restricting $m$ to the valence and conduction bands and using the on-shell condition $E_c - E_v = 2\hbar\omega$. In this case, the sequential contribution reduces to a closed form in terms of band velocities, yielding
\begin{align}
    M^{ab}(\mathbf{k})
    =
    \frac{\Delta v^a\, v_{cv}^b + \Delta v^b\, v_{cv}^a}{\omega}
    +
    w_{cv}^{ab},
\end{align}
where $v_{cv}^a=\langle u_c | v^a | u_v \rangle$ and $\Delta v^a = v_c^a - v_v^a$. The first term originates from the energy-denominator structure of the sequential processes after imposing the resonance condition. The second term, $w_{cv}^{ab}=\langle u_c | \partial_a \partial_b H | u_v \rangle$, arises from the direct two-photon transition. Unlike the sequential contribution, this term does not involve an intermediate virtual state and therefore carries no frequency denominator. Physically, it represents the instantaneous response of the band structure curvature to the applied field and must be included on equal footing with the sequential processes to obtain a finite and gauge-consistent transition amplitude.\\
Therefore, the 2PA transition rate can be expressed in various powers of $\omega$ as
\begin{align}\label{2PAT}
|M^{ab}|^2= \frac{|A_{ab}|^2 }{\omega^2}+\frac{2\re\left[A_{ab}\left(w_{cv}^{ab}\right)^*\right]}{\omega}+\left|w_{cv}^{ab}\right|^2.
\end{align}
$|A_{ab}|^2=\left|\Delta v^a v_{cv}^b+\Delta v^b v_{cv}^a\right|^2$ and $\left|w_{cv}^{ab}\right|^2$ are purely sequential and direct terms, respectively, and the middle term proportional to $\mathcal{O}(\omega^{-1})$ is a mixed term.\\
Importantly, in altermagnets, the contact term is not merely a technical correction to the velocity-gauge formulation. Since the altermagnetic order parameter, $J(\vex{k})$, enters the Bloch Hamiltonian through momentum-dependent even-harmonic form factors, the contact term in even-photon absorption carries explicit information about the same anisotropic harmonic structure that characterizes the altermagnetic order. For example, a $d$-wave altermagnetic order parameter contributes naturally to second derivatives of the Hamiltonian, while higher-harmonic orders such as $g$- and $i$-wave orders generate characteristic structures in 4PA and 6PA. Therefore, retaining contact term is essential not only for gauge consistency, but also for faithfully capturing how the symmetry of the altermagnetic order is encoded in the multi-photon optical response. \\
To illustrate the mechanism, we first consider $d$-wave ALM. For the $d_{x^2-y^2}$ order, $J_{x^2-y^2}(\mathbf{k}) = J[\cos(k_x)-\cos(k_y)]$, the mixed contact term vanishes identically, $w_{cv}^{xy}=0$. The two-photon transition amplitude in the crossed-polarization channel therefore reduces to $M^{xy}=A_{xy}/\omega$, yielding $|M^{xy}|^2 = |A_{xy}|^2/\omega^2$. As a result, the two-photon absorption exhibits the characteristic $\mathcal{O}(\omega^{-2})$ scaling, in sharp contrast to the generic behavior in Eq.~\eqref{2PAT}, where a finite contact term produces a frequency-independent contribution at large $\omega$. A similar result holds for the $d_{xy}$ order, $J_{xy}(\mathbf{k}) = J \sin(k_x)\sin(k_y)$. In this case, the contact term vanishes in the rotated polarization channel $M^{x-y,x+y} \equiv M^{xx}-M^{yy}$, corresponding to polarized photons along the two diagonals of the crystal. The transition amplitude in this channel is again purely velocity-mediated, giving $|M^{x-y,x+y}|^2 \propto \omega^{-2}$.\\
Thus, when the polarization tensor of the absorbed photons is aligned with the anisotropy of the altermagnetic order parameter, the contact term is symmetry-forbidden. The resulting suppression of the constant high-frequency contribution leaves a distinctive $1/\omega^2$ scaling of the two-photon absorption matrix element, providing a direct optical signature of altermagnetic symmetry. \\
\blue{\emph{4PA in $g$-wave ALM}}.---Having established the structure of the two-photon matrix element, we now turn to higher-harmonic altermagnetic orders. For a $d$-wave ALM, the leading even harmonic of the order parameter can already be accessed through the second derivatives of the Bloch Hamiltonian and therefore appears naturally in 2PA. By contrast, a $g$-wave altermagnetic order contains a higher even-harmonic momentum structure, so 2PA does not provide a unique symmetry fingerprint of the $g$-wave form factor. To isolate a response that is characteristic of the $g$-wave order, one must instead examine the next relevant even-photon absorption process, namely four-photon absorption, where fourth-order momentum derivatives and higher-order velocity vertices can directly encode the $g$-wave harmonic structure.\\
The 4PA absorption is composed of five processes: four linear vertices ($v^av^bv^cv^d:A^{abcd}$), one quadratic and two linear vertices ($v^av^bw^{cd}:B^{abcbd}$), two quadratic vertices and one linear and one cubic vertex ($w^{ab}w^{cd}+v^au^{bcd}:C^{abcd}$) and a quartic contact vertex ($x^{abcd}$). Therefore, the on-shell 4PA can be written as
\[
M_{\rm sym}^{abcd}
=
\frac{A^{abcd}}{\omega^3}
+
\frac{B^{abcd}}{\omega^2}
+
\frac{C^{abcd}}{\omega}
+
x^{abcd}_{cv}.
\]
We focus on a specific $g$-wave ALM order $J_g(\vex{k})=J\left[(\cos(k_x)-\cos(k_y))^2-(2\sin(k_x)\sin(k_y))^2\right]$. Thus, for the above \(g\)-wave order, the contact vertex has a distinguished null component, \(x^{xyuv}_{cv}=0\), with \(u=x-y\) and \(v=x+y\). This vanishing follows from the directional-derivative structure of the \(g\)-wave form factor and shows that multiphoton absorption can select polarization tensors matched to the anisotropic momentum-space texture of
the altermagnetic order. As a result, $|M_{\rm sym}^{xyuv}|^2$ also show distinct frequency scaling as well, eliminating $\omega^{0}$ and $\omega^{-1}$ terms, and the response terminates at $\omega^{-2}$. This absence of the \(\omega^{-1}\) and \(\omega^{0}\) contributions provides a direct frequency-domain signature of the null contact vertex selected by the \(g\)-wave altermagnetic form factor. Furthermore, the $\omega^{-2}$ and $\omega^{-3}$ contributions also simplify further due to vanishing $\re[A^{xyuv}(x^{xyuv})^*]$ and $\re[B^{xyuv}(x^{xyuv})^*]$ interference terms. So, we obtain,
\begin{align}
|M_{\rm sym}^{xyuv}|^2
&=
\frac{|A^{xyuv}|^2}{\omega^6}
+
\frac{2}{\omega^5}
\operatorname{Re}\!\left[
A^{xyuv}\left(B^{xyuv}\right)^*
\right]
\cr
&+
\frac{
|B^{xyuv}|^2
+
2\operatorname{Re}\!\left[
A^{xyuv}\left(C^{xyuv}\right)^*
\right]
}{\omega^4}
\cr
&+
\frac{
2\operatorname{Re}\!\left[
B^{xyuv}\left(C^{xyuv}\right)^*
\right]
}{\omega^3}
+
\frac{
|C^{xyuv}|^2
}{\omega^2},
\end{align}
where $|M_{\rm sym}^{xyuv}|^2=|M_{\rm sym}^{xxxy}|^2-|M_{\rm sym}^{xyyy}|^2$. The explicit forms of $A^{xyuv},\,B^{xyuv}$ and $C^{xyuv}$ are provided in the End Matter. \\
It is important to make a distinction from unique 4PA component for $g$-wave and
corresponding mixed channels in a \(d\)-wave ALM which also vanish. For the
\(d\)-wave form factor, $J(\vex{k})_{x^2-y^2}$, due to $w^{xy}=0$ all the mixed quartic contact vertices in 4PA is inherited from a lower-order
constraint: \(x^{xxxy}_{cv}=x^{xyyy}_{cv}=x^{xyxy}_{cv}=0\).  In contrast, the \(g\)-wave form factor considered here has the individual mixed 4PA components generally retain quartic-contact contributions. The null condition is instead restricted to the $x^{xyuv}$
while \(x^{xxxy}_{cv}\), \(x^{xyyy}_{cv}\), and \(x^{xyxy}_{cv}\)
need not vanish separately. Comparing this response with neighboring mixed channels, such as \(|M^{xxxy}_{\rm sym}|^2\), \(|M^{xyyy}_{\rm sym}|^2\), or \(|M^{xyxy}_{\rm sym}|^2\), therefore separates a genuine four-photon \(g\)-wave fingerprint from the broader mixed-channel suppression inherited from \(w^{xy}=0\) in the \(d\)-wave case.

\blue{\emph{6PA in $i$-wave ALM}}.---The same hierarchy extends to still higher
altermagnetic harmonics. An \(i\)-wave altermagnetic order carries a sixth-order
angular structure, and therefore lower-order processes such as 2PA and 4PA do
not provide a unique tensor fingerprint of the \(i\)-wave form factor. The
natural nonlinear probe is instead six-photon absorption, where contact vertex vanishes and can directly encode the \(i\)-wave harmonic structure.\\
The 6PA matrix element contains all vertex partitions of six absorbed photons.
These can be grouped according to their on-shell frequency dependence: six
linear vertices \((v^6:A^{abcdef})/\omega^5\), one quadratic and four linear vertices
\((w v^4:B^{abcdef})/\omega^4\), two quadratic vertices with two linear vertices together
with one cubic vertex and three linear vertices
\((w^2v^2+uv^3:C^{abcdef})/\omega^3\), three quadratic vertices, one cubic-one
quadratic-one linear vertex, and one quartic vertex with two linear vertices
\((w^3+uwv+xv^2:D^{abcdef})/\omega^2\), two cubic vertices, one quartic and one
quadratic vertex, and one quintic vertex with one linear vertex
\((u^2+xw+yv:E^{abcdef})/\omega\), followed by the sixth-order contact vertex
\((z^{abcdef}_{cv})\). Here, we skip the full expressions for 6PA as it includes many terms after symmetrization. However, we show that for a 2D \(i\)-wave altermagnetic order parameter on a triangular lattice, \(J_i(\vex{k})=J d_{2}[3d^2_1-d^2_2]\), where \(d_1=\cos x - \cos\!\left(\frac{x}{2}\right)\cos\!\left(\frac{\sqrt{3}\,y}{2}\right)\), \(d_2=\sqrt{3} \sin{\frac{x}{2}} \sin{\frac{\sqrt{3}y}{2}}\), the unique null contact term is \(z^{\rho_+\rho_-\mu_+\mu_-\nu_+\nu_-}_{cv}\) where $\rho_{\pm}=3\sqrt{3}x\pm y$, $\mu_{\pm}=\sqrt{3}x\pm2y$ and $\nu_{\pm}=\sqrt{3}x\pm5y$, providing a frequency-domain diagnostic of the \(i\)-wave altermagnetic texture.\\
\blue{\emph{Discussion}}.--- We have demonstrated that conventional multiphoton transition measurements \cite{he2008multiphoton} can directly probe ALMs. Henceforth, without loss of generality, we focus on the $d$-wave and make a few brief remarks on other possible measures that could consequently be probed. Due to $C_4\mathcal{T}$ symmetry, both circular and linear dichroisms vanish even for the $2$PA case. However, due to the distinct frequency scaling of the $M^{xy}$ demonstrated in this work, a \emph{crossed linear dichroism} $|M^{xy}|^2-|M^{xx/yy}|^2$ or consequently a two-photon counterpart of a more conventional circular-linear dichroism \cite{} will reflect the hallmark of altermagnetism in $d$-wave ALMs. \\
So far, we focused on degenerate MPA where all photons share the same frequency. However, non-degenerate MPA (where multiple photons carry different frequency) has also been center of interest. Related to 2PA results in ALM obtained in this work, ND-2PA provides an extra feature that would be useful for probing symmetry of ALM. We obtain the ND-2PA $|M^{xy}_{ND}|^2$ for a $d$-wave ALM involving two photons with $\omega$ and $\alpha\omega$ is given as \begin{align}
|M^{xy}_{ND}|^2=&\frac{(\Delta v^x)^2 |v_{cv}^y|^2}{\omega^2}
+
\frac{(\Delta v^y)^2 |v_{cv}^x|^2}{\alpha^2 \omega^2}\cr
&+
\frac{2\Delta v^x \Delta v^y}{\alpha \omega^2}
\operatorname{Re}
\left[
v_{cv}^y (v_{cv}^x)^*
\right].
\end{align}
Interestingly, as a result of $w^{xy}=0$ in the $d$-wave ALM due to polarization locked channel to the symmetry harmonic of the altermagnetic texture, the ND-2PA now involves two symmetry related contributions $(\Delta v^x)^2 |v_{cv}^y|^2$ and $(\Delta v^y)^2 |v_{cv}^x|^2$ that can be suppressed/enhanced relative to each other by variation of relative ratio of two-photon frequencies. Due to underlying  $C_4\mathcal{T}$ symmetry, $\int_{BZ}(\Delta v^x)^2 |v_{cv}^y|^2=\int_{BZ}(\Delta v^y)^2 |v_{cv}^x|^2$, therefore, performing measurement in this polarization channel for two limits of $\alpha\gg1$ and $\alpha\ll 1$ can probe $C_4\mathcal{T}$ breaking in $d$-wave ALM. \\
Additionally, we note that the symmetry fingerprints identified here for multi-photon absorption
should also affect the inverse emission processes \cite{hayat2011applications}. Since the same
velocity and contact vertices enter the nonlinear transition amplitudes,
the even-harmonic momentum structure of the altermagnetic spin splitting
can imprint itself on the polarization, helicity, and harmonic content of the emitted radiation. Thus, the contact vertices that distinguish the $d$-, $g$-, and $i$-wave absorption channels are expected to generate corresponding selection rules in second-, fourth-, and sixth-order emission processes. Multi-photon emission, including high-harmonic generation, can therefore provide a complementary optical probe of altermagnetic order.\\
Finally, the general physical arguments discussed here can be straightforwardly extended to odd-parity magnets~\cite{pwave1,pwave2,song2025electrical,yamada2025metallic}.  In that case, the symmetry-selected null conditions arise in odd-photon absorption processes, reflecting the odd parity of the underlying magnetic form factor. However, for the most widely studied member of this family, namely $p$-wave magnets, the leading harmonic is already
linear in momentum. Therefore, 1PA is sufficient to access the primary \(p\)-wave character, and higher odd-photon processes are not required.

In summary, our results establish multiphoton absorption as a symmetry-selective optical probe of altermagnetic order. By matching the photon order and polarization geometry to the momentum-space spin-splitting harmonic, the response reveals distinct frequency scalings for different altermagnetic textures. This enables \(d\)-, \(g\)-, $i$--wave altermagnets to be distinguished through polarization-resolved absorption measurements. The proposed signatures are directly accessible with nonlinear optical spectroscopy, providing a route toward unambiguous characterization of altermagnetic band symmetry.

\blue{\emph{Acknowledgment}}.--- S.A.A.G was supported by a seed funding from the LRSM at university of Pennsylvania. A.M.R was supported by the U.S. Department of Energy, Office of Science, Basic Energy Sciences, under Award No. DE-SC0024942.

\bibliography{main}

\section{End Matter}
The degenerate four-photon absorption (4PA) rate is written as
\begin{equation}
\Gamma^{(4)}(\omega)
\propto
\int_{-\pi}^{\pi}\!\! dk_x
\int_{-\pi}^{\pi}\!\! dk_y\,
\left|
M_{cv,\mathrm{sym}}^{abcd}(\mathbf k,\omega)
\right|^2
\delta_\eta\!\left(E_{cv}(\mathbf k)-4\omega\right).
\end{equation}
Here \(M_{cv,\mathrm{sym}}^{abcd}\) is the fully symmetrized
four-photon transition amplitude. In the expressions below, we suppress the
explicit \(\mathbf k\)-dependence and use
\begin{equation}
E_{cv}\equiv E_{cv}(\mathbf k),\qquad
v_{mn}^{a}\equiv v_{mn}^{a}(\mathbf k),\qquad
w_{mn}^{ab}\equiv w_{mn}^{ab}(\mathbf k),
\end{equation}
\begin{equation}
u_{mn}^{abc}\equiv u_{mn}^{abc}(\mathbf k),\qquad
x_{mn}^{abcd}\equiv x_{mn}^{abcd}(\mathbf k),
\end{equation}
with \(E_{vv}=0\). After imposing the 4PA resonance only inside the
intermediate-state denominators, one may use
\begin{equation}
E_{cv}-3\omega \to \omega,\qquad
E_{cv}-2\omega \to 2\omega,\qquad
E_{cv}-\omega \to 3\omega .
\end{equation}

For a fixed ordering of the external Cartesian indices \((a,b,c,d)\), we
define the ordered block
\begin{equation}
T^{abcd}
=
T_{vvvv}^{abcd}
+
T_{wvv}^{abcd}
+
T_{ww}^{abcd}
+
T_{uv}^{abcd}
+
\frac{1}{24}x_{cv}^{abcd}.
\end{equation}
The first term contains four linear velocity vertices:
\begin{align}
T_{vvvv}^{abcd}
&=
-\frac{1}{6\omega^3}
v_{cv}^{a}v_{vv}^{b}v_{vv}^{c}v_{vv}^{d}
+
\frac{1}{2\omega^3}
v_{cc}^{a}v_{cv}^{b}v_{vv}^{c}v_{vv}^{d}
\nonumber\\
&\quad
+
\frac{1}{6\omega^3}
v_{cv}^{a}v_{vc}^{b}v_{cv}^{c}v_{vv}^{d}
+
\frac{1}{18\omega^3}
v_{cv}^{a}v_{vv}^{b}v_{vc}^{c}v_{cv}^{d}
\nonumber\\
&\quad
-
\frac{1}{2\omega^3}
v_{cc}^{a}v_{cc}^{b}v_{cv}^{c}v_{vv}^{d}
-
\frac{1}{6\omega^3}
v_{cc}^{a}v_{cv}^{b}v_{vc}^{c}v_{cv}^{d}
\nonumber\\
&\quad
-
\frac{1}{18\omega^3}
v_{cv}^{a}v_{vc}^{b}v_{cc}^{c}v_{cv}^{d}
+
\frac{1}{6\omega^3}
v_{cc}^{a}v_{cc}^{b}v_{cc}^{c}v_{cv}^{d}.
\end{align}
The terms with one quadratic vertex and two linear vertices are
\begin{align}
T_{wvv}^{abcd}
&=
\frac{1}{4\omega^2}
w_{cv}^{ab}v_{vv}^{c}v_{vv}^{d}
-
\frac{1}{4\omega^2}
w_{cc}^{ab}v_{cv}^{c}v_{vv}^{d}
-
\frac{1}{12\omega^2}
w_{cv}^{ab}v_{vc}^{c}v_{cv}^{d}
\nonumber\\
&\quad
+
\frac{1}{12\omega^2}
w_{cc}^{ab}v_{cc}^{c}v_{cv}^{d}
+
\frac{1}{6\omega^2}
v_{cv}^{a}w_{vv}^{bc}v_{vv}^{d}
-
\frac{1}{2\omega^2}
v_{cc}^{a}w_{cv}^{bc}v_{vv}^{d}
\nonumber\\
&\quad
-
\frac{1}{18\omega^2}
v_{cv}^{a}w_{vc}^{bc}v_{cv}^{d}
+
\frac{1}{6\omega^2}
v_{cc}^{a}w_{cc}^{bc}v_{cv}^{d}
+
\frac{1}{12\omega^2}
v_{cv}^{a}v_{vv}^{b}w_{vv}^{cd}
\nonumber\\
&\quad
-
\frac{1}{4\omega^2}
v_{cc}^{a}v_{cv}^{b}w_{vv}^{cd}
-
\frac{1}{12\omega^2}
v_{cv}^{a}v_{vc}^{b}w_{cv}^{cd}
+
\frac{1}{4\omega^2}
v_{cc}^{a}v_{cc}^{b}w_{cv}^{cd}.
\end{align}
The contribution with two quadratic vertices is
\begin{equation}
T_{ww}^{abcd}
=
-\frac{1}{8\omega}
w_{cv}^{ab}w_{vv}^{cd}
+
\frac{1}{8\omega}
w_{cc}^{ab}w_{cv}^{cd}.
\end{equation}
The terms with one cubic vertex and one linear vertex are
\begin{equation}
T_{uv}^{abcd}
=
-\frac{1}{6\omega}
u_{cv}^{abc}v_{vv}^{d}
+
\frac{1}{18\omega}
u_{cc}^{abc}v_{cv}^{d}
-
\frac{1}{18\omega}
v_{cv}^{a}u_{vv}^{bcd}
+
\frac{1}{6\omega}
v_{cc}^{a}u_{cv}^{bcd}.
\end{equation}

The fully symmetrized 4PA amplitude is obtained by summing over all
permutations of the four external indices. Defining
\begin{equation}
\mathcal S_4[F^{abcd}]
\equiv
\sum_{\pi\in S_4}
F^{\pi(a)\pi(b)\pi(c)\pi(d)},
\end{equation}
we have
\begin{equation}
M_{cv,\mathrm{sym}}^{abcd}
=
\mathcal S_4\!\left[T^{abcd}\right].
\end{equation}
Since the quartic contact vertex is fully symmetric,
\(x_{cv}^{abcd}=x_{cv}^{(abcd)}\), its contribution satisfies
\begin{equation}
\mathcal S_4\!\left[\frac{1}{24}x_{cv}^{abcd}\right]
=
x_{cv}^{abcd}.
\end{equation}
Thus, the symmetrized 4PA amplitude can equivalently be written as
\begin{equation}
M_{cv,\mathrm{sym}}^{abcd}
=
\mathcal S_4\!\left[
T_{vvvv}^{abcd}
+
T_{wvv}^{abcd}
+
T_{ww}^{abcd}
+
T_{uv}^{abcd}
\right]
+
x_{cv}^{abcd}.
\end{equation}

Therefore, the 4PA transition probability can be expressed in terms of the following $1/\omega$ series,
\begin{align}
|M_{\rm sym}^{abcd}|^2
&=
\frac{|A^{abcd}|^2}{\omega^6}
+
\frac{2}{\omega^5}
\operatorname{Re}\!\left[
A^{abcd}\left(B^{abcd}\right)^*
\right]
\cr
&\quad+
\frac{
|B^{abcd}|^2
+
2\operatorname{Re}\!\left[
A^{abcd}\left(C^{abcd}\right)^*
\right]
}{\omega^4}
\cr
&\quad+
\frac{
2\operatorname{Re}\!\left[
B^{abcd}\left(C^{abcd}\right)^*
\right]
+
2\operatorname{Re}\!\left[
A^{abcd}\left(x^{abcd}_{cv}\right)^*
\right]
}{\omega^3}
\cr
&\quad+
\frac{
|C^{abcd}|^2
+
2\operatorname{Re}\!\left[
B^{abcd}\left(x^{abcd}_{cv}\right)^*
\right]
}{\omega^2}
\cr
&\quad+
\frac{2}{\omega}
\operatorname{Re}\!\left[
C^{abcd}\left(x^{abcd}_{cv}\right)^*
\right]
+
|x^{abcd}_{cv}|^2 .
\end{align}

\end{document}